\documentclass[aps,pra,10pt,twocolumn,amsmath,amssymb,notitlepage,longbibliography]{revtex4-1}
\usepackage{caption}
\usepackage{subcaption}
\usepackage{tikz}
\usepackage{graphicx}

\usepackage{graphicx, amsmath, amssymb, amsthm, xcolor, bm, bbm, pgfplots,dsfont}

\usepackage{times} 

\pgfplotsset{compat=newest}
\usepgfplotslibrary{external}
\usepackage{soul}
\theoremstyle{plain}

\theoremstyle{definition}

\theoremstyle{remark}

\definecolor{nblue}{rgb}{0.2,0.2,0.7}
\definecolor{ngreen}{rgb}{0.1,0.5,0.1}
\definecolor{nred}{rgb}{0.8,0.2,0.2}
\definecolor{nblack}{rgb}{0,0,0}

\newcommand{\bs}[1]{\boldsymbol{#1}}

\newcommand{\E}{\mathcal{E}}

\newcommand{\unit}{\mathbbm{1}}

\newcommand{\mc}[1]{\mathcal{#1}}

\newcommand{\bc}[1]{\boldsymbol{\mathcal{#1}}}

\newcommand{\density}[1]{|#1\rangle\langle #1|}

\newcommand{\ket}[1]{|#1\rangle}

\newcommand{\bpa}[1]{\bigl(#1\bigr)}

\newcommand{\hidden}[1]{}

\newcommand{\mat}[2]{\left(\begin{array}{#1} #2 \end{array}\right)}

\usepackage{hyperref}
\definecolor{darkblue}{RGB}{0,0,127} 
\definecolor{darkgreen}{RGB}{0,150,0}
\hypersetup{breaklinks, colorlinks, linkcolor=darkblue, citecolor=darkgreen, filecolor=red, urlcolor=darkblue}
\hypersetup{pdftitle={Estimating the fidelity of T gates using standard interleaved randomized benchmarking}, pdfauthor={Robin Harper and Steven T. Flammia}}

\def\equationautorefname~#1\null{Eq.~(#1)\null}

\begin{document}


\title{Estimating the fidelity of \texorpdfstring{$T$}{T} gates using standard interleaved randomized benchmarking}
\author{Robin Harper}
\author{Steven T.\ Flammia}
\affiliation{Centre for Engineered Quantum Systems, School of Physics, The 
University of Sydney, Sydney, Australia}
\date{\today}

\begin{abstract}
Randomized benchmarking (RB) is an important protocol for robustly characterizing the error rates of quantum gates. 
The technique is typically applied to the Clifford gates since they form a group that satisfies a convenient technical condition of forming a unitary 2-design, in addition to having a tight connection to fault-tolerant quantum computing and an efficient classical simulation. 
In order to achieve universal quantum computing one must add at least one additional gate such as the $T$~gate (also known as the $\pi$/8 gate). 
Here we propose and analyze a simple variation of the standard interleaved RB protocol that can accurately estimate the average fidelity of the $T$~gate while retaining the many advantages of a unitary 2-design and the fidelity guarantees that such a design delivers, as well as the efficient classical simulation property of the Clifford group. 
Our work complements prior methods that have succeeded in estimating $T$~gate fidelities, but only by relaxing the 2-design constraint and using a more complicated data analysis. 
\end{abstract}

\maketitle

Randomized benchmarking \cite{Emerson2005, Knill2008, Dankert2009, Magesan2011} uses long sequences of quantum gates with the aim of amplifying small errors in the implementation of the gates, providing a scalable method for quantifying these errors. 
One key advantage of randomized benchmarking (RB) over quantum process tomography is that it is robust to errors associated with state preparation and measurement (SPAM) noise, so that it is able to isolate the contribution of the noise due solely to the gates.
RB is also substantially more scalable than quantum process tomography, and these crucial advantages have made it an indispensable tool for quantum computing experiments.

The Clifford group plays a central role in the theory of RB, and there are several theoretical motivations for this. 
First, it provides a natural class of circuits that can be simulated efficiently on a classical computer. 
Moreover, most fault-tolerant architectures make extensive use of Clifford circuits, so these are precisely the types of gates that will likely appear in implementations. 
The Clifford group also satisfies the technical condition of being a unitary 2-design {\cite{Dankert2009}, meaning that the average over the Clifford group transforms general noise sources into just depolarizing noise. (See \autoref{sec:fidelity} and \cite{Emerson2005}\cite{Nielsen2002} for more details.) 
Finally, standard RB estimates the average error in an ensemble of Clifford gates, and the method of interleaved RB~\cite{Magesan2012, Gaebler2012} enables the estimation of average errors on individual Clifford gates, which yields quite detailed information about the errors in a Clifford circuit given the low cost of the method. 

It is important to be able to estimate the fidelity of the physical realization of non-Clifford gates as well because while Clifford gates play an important role in current approaches to fault-tolerant quantum computing they do not allow for universal quantum computation. 
One must add at least one non-Clifford gate to achieve universality, and the most common choice of additional gate is the so-called $T$~gate, or $\pi/8$ gate. 

Several protocols in the literature have expanded the notion of RB so that it can estimate the average error of $T$~gates. 
Although the unitary 2-design property mentioned above is needed to average out generic errors in the gates, recent work by \citet{Dugas2015} and \citet{Cross2015} has shown that something nearly as good can be obtained by relaxing this condition. 
Instead of the Clifford group, they use a dihedral group for single qubits or a CNOT-dihedral group for multiple qubits and demonstrate a practical method to benchmark $T$~gates (and other gates not in the Clifford group) using this approach. 

Since these protocols involve averages over gate sets that are not unitary 2-designs, there are several points that distinguish them from standard RB or interleaved RB.
The dihedral-type groups are not sufficient to completely average a generic error into a depolarizing channel 
and require fitting to two exponential decay parameters, although mechanisms exist to allow the extraction of the individual decay parameters by analyzing the differences between the average survival probabilities of differently constructed runs. In Ref.~\cite{Kimmel2014} a method of using Randomized Benchmarking to calculate the fidelity of gates by measuring the overlap between a specific Clifford Channel and any target unitary was introduced. This was the first method that demonstrated  the idea of using Randomized Benchmarking to benchmark any unitary and consequently is more flexible than the protocol described in this paper. It alters the Randomized Bencharking protocol slightly to change the channels generated to channels representing specific Clifford `maps' and allows the overlap between these Clifford maps and the target gate to be measured and thus the target unitary to be reconstructed. However, for a single qubit $T$~gate it requires the fidelity of the overlap of three composed channels to the $T$~ gate to be estimated and then these overlaps are `best-fit' to obtain an estimated $T$~gate fidelity. In addition the overlap between the specific Clifford channels and the $T$~gate will not be near unity, increasing the binomial variance of the results --- meaning many more samples are required for any specific accuracy.

In Ref.~\cite{Barends2014} two $T$~gates were inserted between each Clifford element, so that the fidelity of a double application of a $T$~gate could be estimated. 
One can then extract the average error on a single $T$~gate if one is willing to assume that the errors are not correlated 
and compose in a straightforward manner. 
This scheme has the advantage of being a unitary-2 design, but compounds the estimation errors that occur when attempting to decompose error sources from the estimation of the composed errors. Finally in Ref.~\cite{Chasseur} a protocol is presented that obviates the need to calculate an inversion gate at the end of the sequence by combining RB with the use of Monte Carlo sampling for fidelity estimation~\cite{Flammia2011, da-Silva2011}. While such a protocol allows for fidelity estimation for arbitrary quantum gates it scales exponentially in the number of qubits. 

Here we propose and analyze a simple variation of the standard interleaved RB protocol that, under the usual RB assumptions explained below, at least diagonalizes the noise between subsequent $T$~gates. 
This renders the protocol more robust to coherent errors on the $T$~gate. 
As a consequence this extension is a method of measuring the average fidelity of a $T$~gate in a scalable fashion, while preserving the advantageous unitary 2-design property and using only elements in the Clifford group (except for the interleaved $T$~gates). 
In fact, our method works on a slightly more general set of gates that includes $T$ as a special case, as we discuss below.
Finally, this approach provides a simplification with respect to the fit parameters and, within the assumptions made, retains the strong theoretical guarantees on the error associated with RB \cite{Wallman2014, Kueng2016, Ball, Wallman2015}. 

\section{Protocol}\label{sec:protocol}

Randomized benchmarking uses the Clifford group (or any unitary 2-design) to efficiently estimate the average fidelity of a noise map $\mathcal{E}$ associated with this group of operations. 
Here we are implicitly assuming that this noise map obeys various convenient simplifying assumptions, namely that the noise is time-independent and gate-independent, such that the noise really is characterized by a single map $\mathcal{E}$.
Below we will discuss what happens when these simplifying assumptions are relaxed.
The average fidelity of $\mathcal{E}$ is defined as
\begin{equation}\label{eq:avgF}
\mathcal{F}_\text{avg}(\mathcal{E}):=\int{\rm d}\psi \langle \psi | \mathcal{E}(\psi) | \psi\rangle\,,
\end{equation}
where the average is over the unitarily invariant Haar measure. 
The average fidelity of a noise map can be efficiently estimated by twirling the error operator over the Clifford group (or indeed any unitary 2-design). 
The twirling operation over a group $\mathcal{G}$ is defined as simply performing the group average, yielding
\begin{align}\label{eq:twirl}
	\mathcal{E}^{\mathcal{G}} = \frac{1}{\lvert \mathcal{G}\rvert} \sum_{g \in \mathcal{G}} g^\dag \mathcal{E} g\,,
\end{align} 
where $g$ is the unitary operation corresponding to a group element. 
This operation of twirling preserves the average fidelity of the noise map $\mathcal{E}$, and even more, it outputs a pure depolarizing channel that can be used to estimate the average fidelity by doing a simple fit to an exponential decay curve. 
As alluded to above, the advantages of using the Clifford group in this twirling operation include that Clifford gates are relatively easy to implement and that the gates form a group so that their operation can be efficiently inverted after a random sequence of such operators. 
The partial information provided by the average fidelity is useful in practice for improving the implementation of the gates as well as providing a bound on the threshold error rate required for fault-tolerant quantum computing~\cite{Sanders2015, Kueng2016, Wallman2015}.

The set of gates consisting of any Clifford group element $C$ followed by $T$, a $T$~gate, is a unitary 2-design. 
This is intuitively clear given that the Clifford group itself forms a unitary 2-design, and it is easily verified by calculating the frame potential of the gate set \cite{Gross2007a}. 
Consequently, this $TC$ gate set can be used in some related protocols that require a unitary 2-design, such as measuring unitarity \cite{Wallman2015a} or measuring loss rates \cite{Wallman2014a} (which in fact only requires a weaker condition of a unitary 1-design). 
However, this set of gates does not form a finite group. 
In fact its image is dense on the entire unitary group, so there is no scalable method of inverting the sequence of gates at the end of a benchmarking sequence and it becomes difficult to use the gates to measure average fidelity in an accurate and precise way.

We now describe the protocol that allows $T$~gates to be used with the randomized benchmarking protocol while preserving the unitary 2-design property. 
First we note that the $T$~gate is a gate in the third level of the Clifford hierarchy~\cite{Gottesman1999}. 
This means that it conjugates members of the Pauli group to the Clifford group (ignoring any overall global phases), namely for every Pauli gate $P$ there exists a Clifford $C$ such that
\begin{equation}
T P T^\dagger = C\,.
\end{equation}
For an arbitrary number of qubits we will write $\mathcal{C}$ to represent the normal Clifford gates, $\mathcal{P}$ to represent the Pauli operators and $\mathcal{C}_t$ to represent the subset of Clifford gates formed by conjugation of the elements of $\mathcal{P}$ with a $T$~gate.

In this protocol gates from the Clifford group (or a subgroup of the Clifford group, as discussed in Sec.~\ref{sec:cleve}) form the basis of our unitary-2 design. 
We will denote this group of Clifford gates by $\mathcal{G}$, and the full gate set we use will be any gate of the form $TC$ for $C \in \mathcal{G}$. 
In line with standard randomized benchmarking assumptions we assume that an experimental implementation of a gate $U \in \mathcal{G}$ can be written as $\mathcal{U}\mathcal{E}$ where $\mathcal{U}$ denotes the channel corresponding to conjugation by $U$ and $\E$ is a completely positive trace preserving (CPTP) channel independent of $U$.
We also note that the assumption that $\E$ is independent of $U$ can be relaxed without significant effect on the results \cite{Epstein2014,Magesan2011,Wallman2014a}. Specifically since the Pauli gates are a subset of the Clifford group we are assuming the error on the Pauli gates is the same as for the other Clifford gates (subject to the  relaxation noted above).
With respect to the interleaved gate (here the $T$~gate) it is anticipated that the error channel associated with that gate will be different from those forming $\mathcal{G}$, and we will treat this error separately.

Our $T$~gate interleaved RB protocol is very similar to the usual interleaved randomized benchmarking procedure. We begin by determining the reference fidelity using a slight modification of standard benchmarking. 
But first some notation: 
If $\mathcal{G}$ is a group, then $|G|$ denotes the total number of elements in $\mathcal{G}$. We also say, by an abuse of notation, that $|\mathcal{G}|$ is the set $\{1,\ldots,|\mathcal{G}|\}$. Thus, a statement such as $j \in |\mathcal{G}|$ simply means that $j$ is a label for the $j$th element in $\mathcal{G}$.

\begin{figure*}[!htb]
\begin{subfigure}[b]{0.32\textwidth}
	 \includegraphics[width=1\columnwidth]{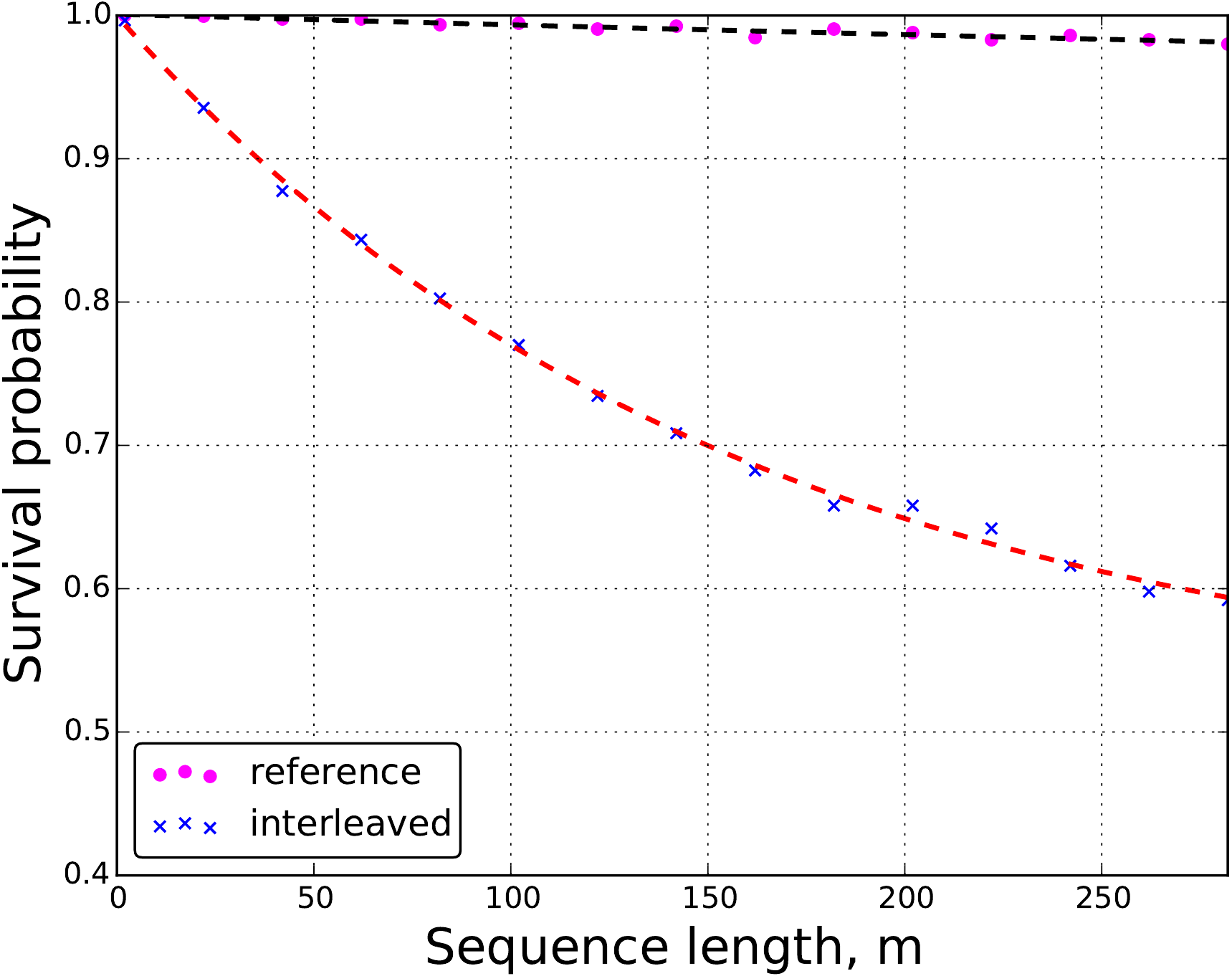}
	 \caption{Rotation Errors}
	 \label{fig:rotation}
\end{subfigure}
\begin{subfigure}[b]{0.32\textwidth}
	 \includegraphics[width=1\columnwidth]{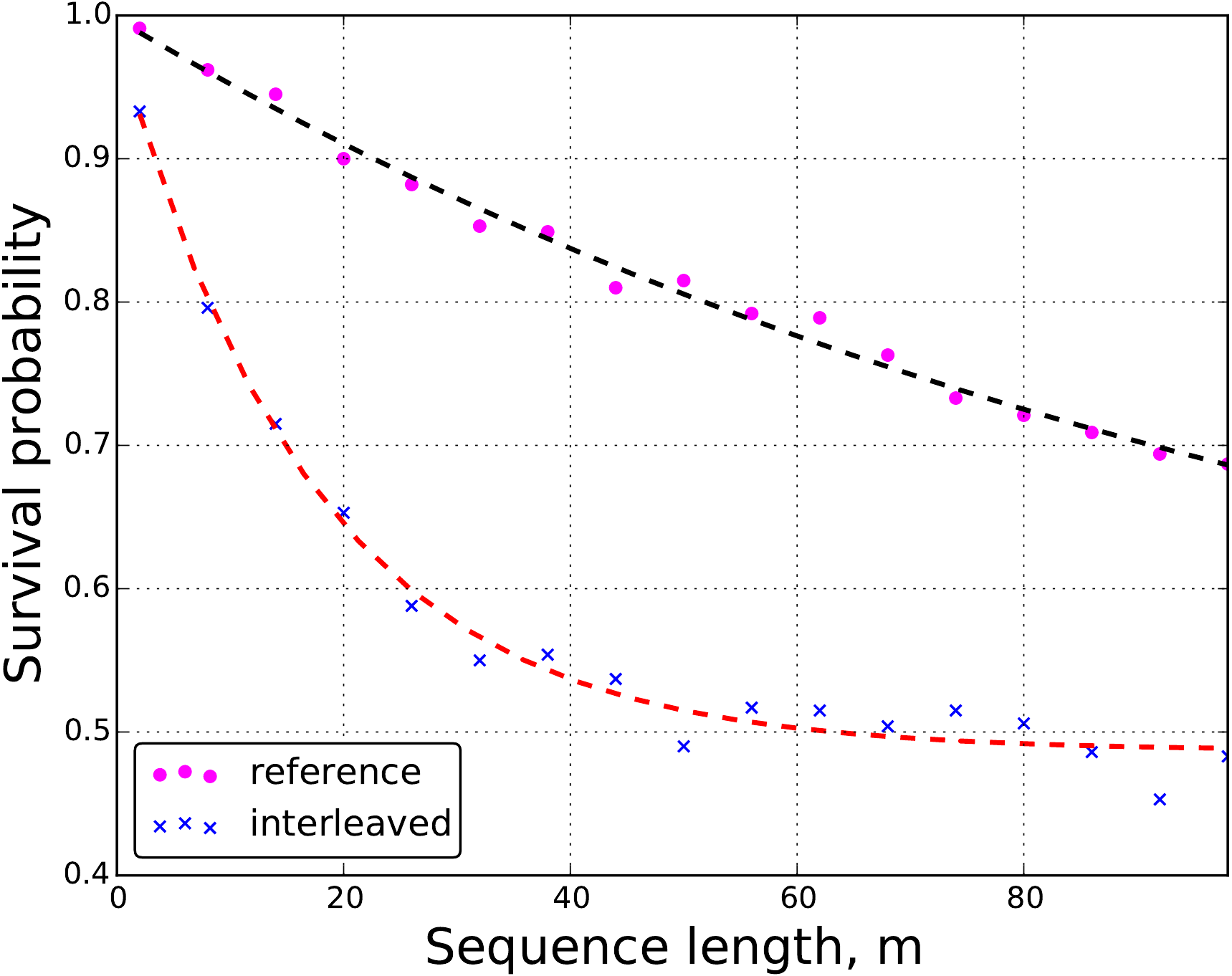}
	 \caption{Depolarizing  and Rotation Errors}
	 \label{fig:depol}
\end{subfigure}
\begin{subfigure}[b]{0.32\textwidth}
	 \includegraphics[width=1\columnwidth]{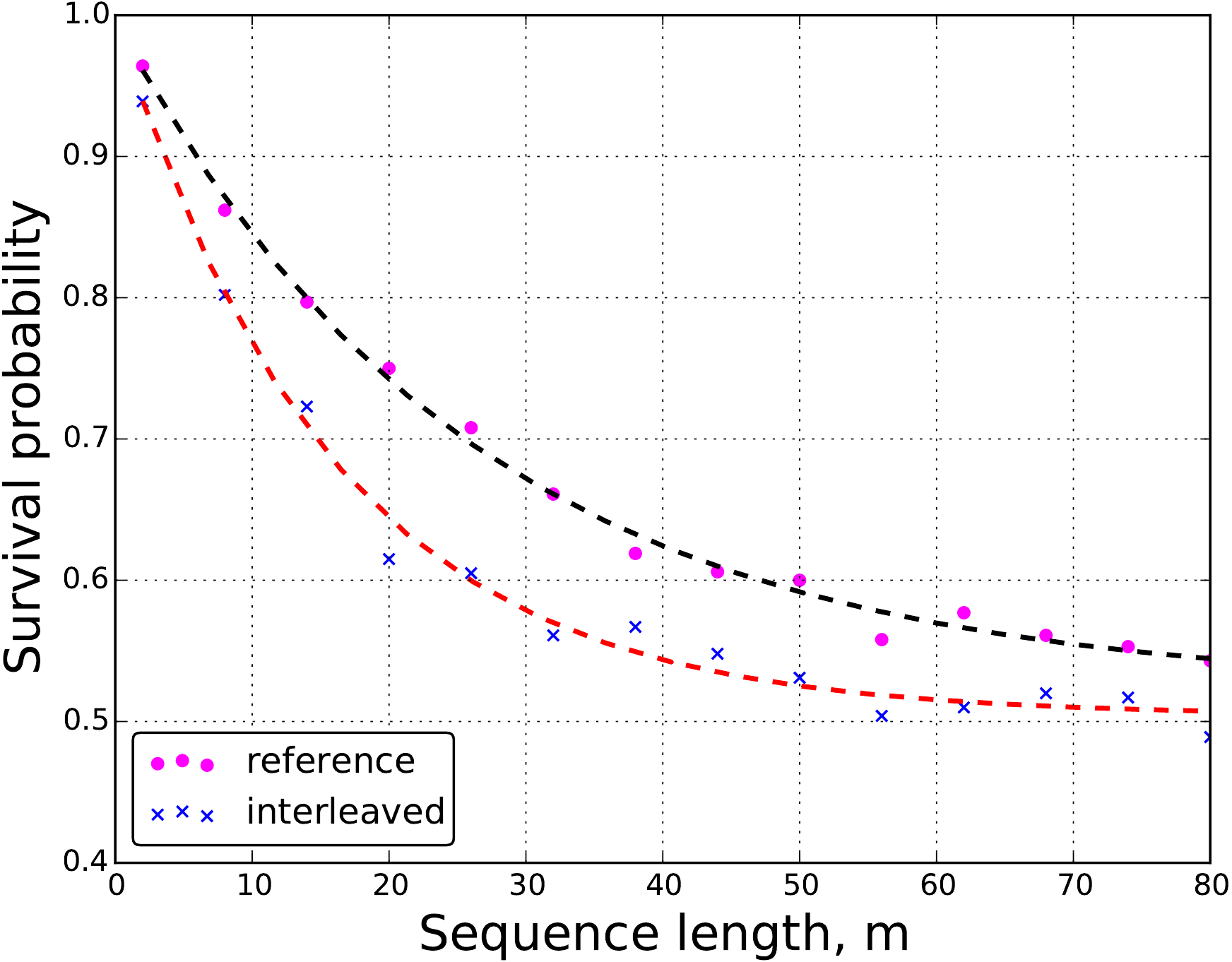}
	 \caption{Amplitude Damping channel}
	 \label{fig:damp}
\end{subfigure}
 \caption{In (a) we simulate a typical randomized benchmarking plot where the gates making up the Clifford group are assumed to suffer from a systematic over-rotation around the x-axis of 0.02 radians (ie $e^{(-0.01iX)}$) , whereas the $T$~gate suffers an over-rotation of 0.12 radians. 
The fidelity of the Clifford gates is 99.993\%. 
In this run the estimated fidelity of the $T$~gate is 99.73\%, compared with the actual fidelity of 99.76\%. 
In this regime the error bounds given by \autoref{Eq:error} are $\pm$0.08\%. In (b) we simulate the application of a depolarizing and rotation error channel. 
The Clifford gates are affected by a depolarizing channel (\autoref{eq:dep}) with $p=0.01$ with an over-rotation of 0.01 radians whereas the $T$~gates are affected by a depolarizing channel with $p=0.02$ and an over-rotation of 0.05 radians. 
The estimated fidelity of the $T$~gates is 97.6\% compared to the actual fidelity of 97.2\%. Finally in (c) a generalized amplitude damping channel has been simulated (see \autoref{eq:amp}). 
For the Clifford group $p=0.995$ and $\gamma=0.01$. 
For the $T$~gate $p=0.99$ and $\gamma=0.04$. 
The Estimated fidelity of the $T$~gate is 98.8\% and the actual fidelity is 98.7\%, for a relative error of about $10^{-3}$.\hspace*{\fill}}
 \label{fig:1}
\end{figure*}

\subsection{Determining the reference fidelity}
\begin{enumerate}
	\item For an even integer $m = 2n$, choose a sequence ${\bf{j}} = (j_1,\ldots,j_n)$ for $j_k \in |\mathcal{G}|$ and a sequence ${\bf{p}} = (p_1,\ldots,p_{n})$ for $p_k \in |\mathcal{P}|$, both uniformly at random.
	\item For each sequence	
	\[U_{\bf{j}}P_{\bf{p}}=U_{j_n}P_{p_n}U_{j_{n-1}}P_{p_{n-1}}\ldots U_{j_1}P_{p_1}\]
	determine the gate $U_{\text{inv}}$ which is $(U_{\bf{j}}P_{\bf{p}})^\dagger$
	\item Apply the sequence 
	\[U_{\text{inv}}U_{j_n}P_{p_n}U_{j_{n-1}}P_{p_{n-1}}\ldots U_{j_1}P_{p_1}\]
	to some initial state {$\rho \neq \frac{1}{d}\unit$ (Usually taken to be the pure state $\ket{0}$) and perform a POVM $\{E,1-E\}$ for some $E$ (Usually taken to be $\density{0}$)}. 
	\item Repeat steps 1 to 3 sufficiently many times to estimate the survival probability to a desired precision over the randomized sequences. 
(See \cite{Granade2014,Wallman2014} for guidance on choosing a sufficient number of samples.)
	\item Repeat steps 1 to 4 for different values of $m$ and fit the results for the averaged sequence fidelity ($\bar{F}_\text{ref}$) to the standard randomized benchmarking model:
	\begin{equation}
	\bar{F}_{\text{ref}}(m) =A_0p_{\text{ref}}^m+B_0
	\end{equation} 
	where $A_0$ and $B_0$ are constants that absorb SPAM errors.
\end{enumerate}
\subsection{Determine the interleaved fidelity}\label{sec:interleaved}
\begin{enumerate}
	\item For an even integer $m=2n$, choose a sequence ${\bf{j}} = (j_1,\ldots,j_n)$ for $j_k \in |\mathcal{G}|$ and a sequence ${\bf{p}} = (p_1,\ldots,p_{n})$ for $p \in |\mathcal{P}|$, both uniformly at random.
	\item For each sequence	
	\[\qquad\quad U_{\bf{j}}TP_{\bf{p}}T=U_{j_n}TP_{p_n}TU_{j_{n-1}}TP_{p_{n-1}}\ldots U_{j_1}TP_{p_1}T\] 
	determine the gate $U_{\text{inv}}$ which is $(U_{\bf{j}}TP_{\bf{p}}T)^\dagger$
	\item Apply the sequence 
	\[U_{\text{inv}}U_{j_n}TP_{p_n}TU_{j_{n-1}}TP_{p_{n-1}}T\ldots U_{j_1}TP_{p_1}T\]
	to the initial state $\rho$ {used in A.3 and measure as chosen in A.3.}
	\item Repeat steps 1 to 3 sufficiently many times to estimate the survival probability to a desired precision over the randomized sequences. 
	\item Repeat steps 1 to 4 for different values of m and fit the results to the model
	\begin{equation}
		\bar{F}_{T}(m) = A_1 p_T^m + B_1 \,.
	\end{equation}
\end{enumerate}

We note that for the purpose of calculating the inverse gate in step 2 of the interleaved protocol that each $P_iT$ can be replaced by an equivalent sequence $TC_i$ where $C_i \in \mathcal{C}_t$. 
The sequence therefore is equivalent to 
\[U_{j_n}TTC_{i_{n}t}U_{j_{n-1}}TTC_{i_{n-1}t}\dots U_{j_1}TTC_{i_1t}\,,\]
which as the product $TT$ gives a Phase gate, collapses to a Clifford circuit, and the inverse gate can then easily be found.

These procedures allow one to collect estimates of $p_{\text{ref}}$ and $p_{T}$. 
The latter quantity conflates the error of the Clifford gates with the interleaved $T$~gates. 
We now wish to relate $p_T$ to the average fidelity of a single $T$~gate, subject of course to some natural assumptions on the noise. In the next section we demonstrate how the fidelity of the $T$ gate can be estimated as 

\begin{equation}
\mathcal{F}(\mc{E}_T)= \frac{d\frac{(d^2-1)p_{T}+1}{(d^2-1)p_{\text{ref}}+1}+1}{d+1}.
\label{eq:tfidelity}
\end{equation}

As an aside, we see that that our protocol works more generally than on just $T$~gates. 
In fact, our protocol works for every gate $A$ such that both $A^2$ and $A P A^\dagger$ are elements of the Clifford group for all Pauli operators $P$. 
This condition is satisfied for many but not all of the square roots of Clifford gates.

\section{Fidelity Estimation}\label{sec:fidelity}

The average fidelity of a general noise map is related to a depolarizing channel (\cite{Emerson2005}\cite{Nielsen2002})
\begin{equation}
\mathcal{D}_p(\rho)=(1-p)\rho+ p\frac{\unit}{d} \,,
\label{eq:dep}
\end{equation}
with a specific depolarizing parameter $p$. 
For a $d$-dimensional quantum system, the depolarizing parameter is related to the average fidelity by
\begin{equation}
\mathcal{F}_\text{avg}(\mathcal{D}_p)=1-p\frac{d-1}{d}\,.
\end{equation}
The connection to arbitrary noise maps $\mathcal{E}$ crucially uses the property of a unitary 2-design. Sampling over a unitary 2-design reproduces the second moments of the Haar measure over all unitaries. As shown in \cite{Emerson2005}\cite{Nielsen2002} this means that if the group $\mathcal{G}$ is a unitary 2-design, we have the following identity  
\begin{equation}
	\mathcal{E}^{\mathcal{G}} = \mathcal{D}_p\,,
\end{equation}
where $p$ is determined by the relations
\begin{equation}
	\mathcal{F}_{\text{avg}}(\mathcal{E}) = \mathcal{F}_{\text{avg}}(\mathcal{E}^{\mathcal{G}}) = \mathcal{F}_{\text{avg}}(\mathcal{D}_p)\,,
\end{equation}
yielding
\begin{equation}\label{eq:pandF}
p=\frac{d}{d-1}\bigl(1-\mathcal{F}_\text{avg}(\mathcal{E})\bigr) \,.
\end{equation}
This is the fundamental result of randomized benchmarking, and relates the average gate fidelity of the noise $\mathcal{E}$ to the fit parameter $p$ of the benchmarking experiments. 
Note that the other crucial assumptions for this derivation to hold are that the noise is time- and gate-independent. 

Our design differs slightly in that it embeds an average over a unitary 1-design (the Pauli group) inside an average over a unitary 2-design (the Clifford group). 
Consequently, while the resulting channel is a depolarizing channel, the depolarizing parameter differs from that given by standard randomized benchmarking by a factor related to the  anisotropy of the noise. 
We now quantify the deviation and show that in the regime of interest this difference is not significant.

To show this it is convenient to work in the superoperator representation of quantum channels. 
The superoperator representation is defined relative to a trace-orthonormal operator basis. 
Here we will use the Pauli basis representation of a channel $\mathcal{E}$, which chooses the suitably normalized Pauli operators as the trace-orthonormal basis. 
This representation consists of a matrix of inner products between the each Pauli matrix ($P_j$) and $\mathcal{E}(P_j)$ (see e.g.~\cite{Wallman2014} for more details). 
Density operators $\rho$ are represented by column vectors $|\rho)$ whose $j$th element is $\text{Tr}(P_j^\dagger \rho)/\sqrt{d}$. 
In particular, it is customary to fix $P_1$ to be the rescaled identity operator~$\unit/\sqrt{d}$. 
In this case any completely positive (CP) channel $\mathcal{E}$ can be written in block form as 
\begin{align}\label{eq:block_terms}
\bs{\E} = \mat{cc}{S(\mc{E})  & \bs{\E}_{\rm sdl} \\ \bs{\E}_{\rm n} 
& \bs{\E}_{\rm u}},
\end{align}
where we refer to $\bc{E}_{\rm sdl}$, $\bs{\E}_{\rm n}$ and 
$\bs{\E}_{\rm u}$ as the \textit{state-dependent leakage}, \textit{nonunital} 
and \textit{unital} blocks respectively (see \cite{Wallman2015a} for more details about this decomposition). 
If the channel is trace preserving then $\bc{E}_{\rm sdl}=0$. 
The unital block ($\bs{\E}_{\rm u}$) contains all the information necessary to extract the fidelity of the channel and, in particular, when the channel is twirled by a unitary-2 design, the resultant noise channel will have $\bs{\E}_{\rm u}$ diagonalized, with each element being the average of the diagonal elements of the original $\bs{\E}_{\rm u}$. 
By way of illustration in the single-qubit case, a completely positive trace preserving noise channel will be of the form:
\begin{align}\label{eq:general_noise}
\bs{\E} = \mat{cccc}{1 & 0 & 0 & 0\\ \alpha_1 &\sigma_x &\delta_1 &\delta_2\\ \alpha_2 &\delta_3 &\sigma_y &\delta_4 \\ \alpha_3  & \delta_5 &\delta_6 & \sigma_z}
\end{align}
where all the elements are real. 
The matrix elements themselves obey certain constraints on account of the requirement for complete positivity; see Ref.~\cite{King2001} for an explicit description of these constraints. 
We will make explicit use of the parameters $\sigma_x, \sigma_y,$ and $\sigma_z$ below.

After averaging over a unitary 2-design, the channel maps to $\mc{E}_\text{twirled} = \frac{1}{|G|}\sum\limits_{C\in G}(C^\dagger \mc{E} C)$, and the unital block $\bs{\E}_{\rm u}$ of the twirled error channel looks like
\begin{align}
\mat{ccc}{\frac{1}{3}({\sigma_x +\sigma_y} + \sigma_z)&0&0\\0&\frac{1}{3}({\sigma_x +\sigma_y} + \sigma_z)&0\\0&0&\frac{1}{3}({\sigma_x +\sigma_y} + \sigma_z)}
\end{align}
where the depolarizing factor measured by randomized benchmarking is $\frac{1}{3}({\sigma_x +\sigma_y} + \sigma_z)$. 
In particular if the twirled noise channel is composed with a twirled noise channel twice then we obtain
\begin{equation}\label{eq:double}
\frac{1}{|G|^2}\sum\limits_{C\in G}(C^\dagger \mc{E} C)\sum\limits_{C\in G}(C^\dagger \mc{E} C)=\left(\frac{1}{|G|}\sum\limits_{C\in G}(C^\dagger \mc{E} C)\right)^2
\end{equation}
and the depolarizing factor is simply squared, giving 
\begin{align}
\label{eq:CCtwirl}
	\bpa{\tfrac{1}{3}({\sigma_x +\sigma_y} + \sigma_z)}^2\,. 
\end{align}
Recalling that $C$ represents a perfect gate taken from the Clifford group and $\mc{E}$ represents the error channel equating to the average error on these gates, this then gives us the depolarizing parameter equating to the average fidelity after the application of two faulty Clifford gates.

In our protocol rather than applying two separate Clifford twirls, in the non-interleaved version, we apply a Clifford gate followed by a Pauli gate. 
Using the techniques described in Ref.~\cite{Magesan2012a} we can decompose a random benchmarking sequence as follows:
\begin{align}\label{eq:initial_decomp}
C_\text{inv} ... \mc{E}  P_2  \mc{E}  C_2   \mc{E}  P_1  \mc{E}  C_1\,.
\end{align}
Where we are using the standard assumptions that $\mc{E}$ is the gate error channel for the gates in the Clifford group (including the Paulis), and have absorbed the error on the inverting gate into the SPAM.  
We can always rewrite $C_2$ as $\tilde{C_2} C_1^\dagger P_1^\dagger $, which means that \autoref{eq:initial_decomp} becomes:
\begin{align}
C_\text{inv} ... \mc{E}  P_2  \mc{E} \tilde{C_2} \Bigl[C_1^\dagger \Bigl[P_1^\dagger  \mc{E}  P_1\Bigr]  \mc{E}  C_1\Bigr]\,.
\end{align}
When the channel is averaged over multiple runs with random choices for each of the Cliffords and Paulis, then for an $n$-gate sequence of Pauli gates followed by Clifford gates, we obtain
\begin{align}
C_\text{inv} \mc{E} \Bigl[C_1^\dagger \Bigl[P_1^\dagger  \mc{E}  P_1\Bigr]^n  \mc{E}  C_1\Bigr]^n
\end{align}
which equates to a depolarizing channel 
\begin{align}\frac{1}{|G_C|}\sum\limits_{C\in G_C}\left(C^\dagger \mc{E} \left(\frac{1}{|G_P|}\sum\limits_{P\in G_P}P^\dagger \mc{E} P\right) C\right)
\end{align} where $G_C$ represents the Clifford group and $G_P$ the Pauli group. In superoperator form in the Pauli basis, the Clifford-Pauli twirled noise of \autoref{eq:general_noise} is of the form:
\begin{align}\label{eq:pc}
\mat{ccc}{\frac{1}{3}(\sigma_x^2 +\sigma_y^2 + \sigma_z^2)&0&0\\0&\frac{1}{3}(\sigma_x^2 +\sigma_y^2 + \sigma_z^2)&0\\0&0&\frac{1}{3}(\sigma_x^2 +\sigma_y^2 + \sigma_z^2)}
\end{align}
which yields a depolarizing parameter of 
\begin{align}
\label{eq:CPtwirl}
	\tfrac{1}{3}\bigl(\sigma_x^2 +\sigma_y^2 + \sigma_z^2\bigr)\,.
\end{align}

In summary, two applications of a Clifford twirl yields the square of the average of the $\sigma_j$ noise parameters (\autoref{eq:CCtwirl}), while the Clifford-Pauli twirl yields the average of the square of these parameters (\autoref{eq:CPtwirl}).

The difference between \autoref{eq:CPtwirl} and \autoref{eq:CCtwirl} can be viewed as the 
\emph{variance} in the depolarizing parameters. 
For the purposes of this protocol we are assuming that the experimental Clifford gates are reasonably high fidelity. 
This means that it is reasonable to assume that the fidelity parameters are individually high, since for small dimensions $d$ the average cannot be high unless each term is reasonably high. 
In this regime, the variance must also be low, as can be seen from an analysis of the anisotropy in the noise. 
Let us write the diagonal elements of $\bc{E}_u$ as being 
\begin{align}
	\sigma_x = \sigma \ , \quad \sigma_y = \sigma (1 + \epsilon_y)\ , \quad \sigma_z = \sigma (1 + \epsilon_z)\,,
\end{align} 
where $\sigma$ is close to $1$ and $\epsilon_y, \epsilon_z \le \epsilon$ are small. 
Expanding the difference of \autoref{eq:CPtwirl} and \autoref{eq:CCtwirl} as a function of the anisotropy of the noise, it can be seen that the variance cancels to $\mc{O}(\sigma^2 \epsilon^2)$. 
Accordingly in the regime of reasonably accurate gates, the error introduced by the additional Pauli twirl is not significant in calculating the average fidelity of the gates. 
Our numerical analysis confirms this (see especially \autoref{fig:channels}).

For completeness we note that when $T$~gates are introduced the protocol can be written as 
\begin{align}
C_\text{inv} \mc{E} ... T  \mc{E}_t  \mc{E}  P_1    T  \mc{E}_t  \mc{E}  C_1 \,,
\end{align}
where the error on the $T$~gate has been written as $\mc{E}_t$, so the noisy $T$~gate is $T  \mc{E}_t$. 
Using the method outlined in \cite{Magesan2012}, this can be rewritten as:
\begin{align}
C_\text{inv} \bigg[C  \mc{E}_t  \mc{E}  \bigg[(PT)^\dagger    \mc{E}_t \mc{E} (PT)\bigg]^n  C\bigg]^n \,.
\end{align}
Noting that $PT$ (being a perfect Pauli followed by a perfect $T$~gate) is a unitary 1-design, it can be seen that the decay parameter given by the protocol is that of the composed error channel $\mc{E}_t  \mc{E}$ and (within the error noted above) gives us the fidelity of the composed error channel.

Now we wish to extract the specific contributions to the noise from the interleaved $T$~gate.
As above, we will denote this noise by $\mathcal{E}_t$. 
The average gate fidelity of \autoref{eq:avgF} of the gates forming the group $G$, being $\mathcal{F}_\text{avg}(\mathcal{E})$ and the average gate fidelity of the the combined $\mathcal{G}T$ gates, being $\mathcal{F}_\text{avg}(\mathcal{E}_t\mathcal{E})$ by using the value of $p$ calculated in step 5 of the relevant part of the protocol (where $p$ is $1-p_{\text{ref}} \text{ and } 1-p_{T}$ respectively) and using \autoref{eq:pandF}. 
The average fidelity of the $T$~gate, $\mathcal{F}_\text{avg}(\mathcal{E}_t)$ can then be estimated from the approximation $\chi_{00}^{ \mathcal{E}_t \mathcal{E}} = \chi_{00}^{\mathcal{E}_t}\chi_{00}^{\mathcal{E}}$, where for qubits 
 \begin{equation}
 \chi_{00}^{\mathcal{E}}=\frac{3}{2}\mathcal{F}_\text{avg}(\mathcal{E})-\frac{1}{2}\,,
 \end{equation}
where here $\chi_{00}^{\mathcal{E}}$ represents the upper left entry of the $\chi$ matrix representation of $\mathcal{E}$ \cite{Nielsen2011}. It is convenient to write the estimated $T$~gate fidelity in terms of the measured depolarising parameters; \cite{Arnaud2016} has a useful table showing how $\mc{F},p,r,\text{ and }\chi_{00}$ are related and this allows us to write the estimate fidelity of the $T$~gate as  \autoref{eq:tfidelity}.
The above approximation is valid to within the bound derived in \cite{Kimmel2014}:
 \begin{multline}
 \chi_{00}^{\mathcal{E}_t\mathcal{E}}-\chi_{00}^{\mathcal{E}_t}\chi_{00}^{\mathcal{E}}\leq2
 \sqrt{(1-\chi_{00}^{\mathcal{E}})\chi_{00}^{\mathcal{E}}(1-\chi_{00}^{\mathcal{E}_t})\chi_{00}^{\mathcal{E}_t}}\\ 
 + (1-\chi_{00}^{\mathcal{E}})(1-\chi_{00}^{\mathcal{E}_t})\,.
 \label{Eq:error}
 \end{multline}
As noted in \cite{Dugas2015} this bound is loose in general but tight in the regime where the gates forming $G$ have a much higher fidelity than the $T$~gate, which is fortunately the regime of interest.

\section{Numerical Simulation} \label{sec:bounds}

\begin{figure}
 \centering
 \includegraphics[width=\columnwidth]{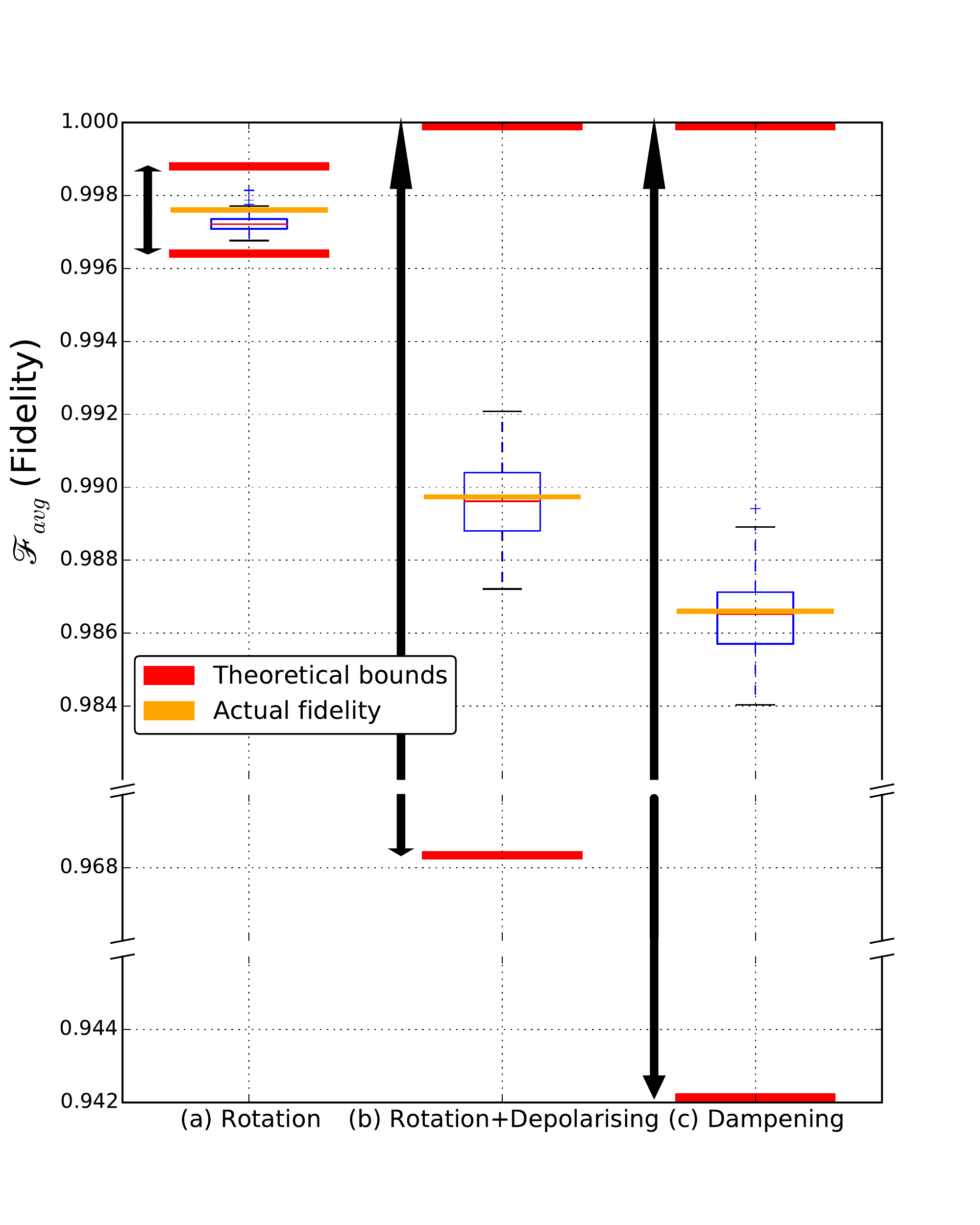}
 \caption{Tukey plots showing estimated fidelities over 100 runs of each of the scenarios presented in \autoref{fig:1}. The box represents the inter-quartile range (IQR), the whiskers 1.5 times the IQR, with outliers plotted separately. The theoretical bounds are shown as thick red lines and the actual fidelity of the $T$~gate as a longer orange line. As can be seen in the regime of \autoref{fig:rotation} the theoretical bounds of \autoref{Eq:error} are reasonably tight, but in the other cases are a lot looser than the simulated results.\hspace*{\fill}}
 \label{fig:2}
\end{figure}

As previously discussed randomized benchmarking is robust, both in theory and practice, to some level of gate dependent noise \cite{Magesan2012,Wallman2014}. 
We now present some numerical simulations illustrating that the robustness in the protocol presented where (as anticipated) the error profile of the $T$~gate is different from the gates in $\mathcal{G}$. 
For the purpose of these simulations we are assuming a single qubit, where $\mathcal{G}$ is taken to be all 24 single-qubit Cliffords $\mathcal{C}$ and $\mathcal{P}$ is the four single-qubit Paulis (which are contained in $\mathcal{C}$). 

Initially, three different noise models were explored. 
As with all interleaved protocols care must be taken to reduce the error of each of the estimates, as effectively the protocol requires one estimation to be divided by another. 
In each simulation, for various gate lengths, a number of random sequences (2000 for \autoref{fig:rotation} and 1000 for \autoref{fig:depol} and \autoref{fig:damp}) were generated. 
At the end of each sequence a single measurement (returning 1 or 0) was made. 
 This follows the design outlined in \cite{Granade2014}. 
Curves were fit using standard least square methodologies. 
\autoref{fig:rotation} illustrates the case where the rotational errors on the Clifford gates are different from that of the implementation of the $T$~gate (modelled as over-rotatons on the x-axis ie $e^{(-iX\frac{\theta}{2})}$ for some $\theta$ -- see caption). 
As previously discussed the low error rate on the Clifford gates allows an accurate estimation of the $T$~gate even with rotational errors. 
The first plot in  \autoref{fig:2} shows the variance in the estimation over 100 simulations as a Tukey box plot. 
As can be seen even in this region, the bounds given by \autoref{Eq:error} are not tight, although it can be seen the protocol does, slightly, underestimate the fidelity of the $T$~gates in this example. 
The 100 simulations have a median error estimate of 99.72\% with a $\sigma$ of 0.00025 (0.025\%), compared with an actual fidelity of 99.76\%.

\begin{figure*}[t!]
\begin{subfigure}[t]{\columnwidth}
\includegraphics[width=1.1\columnwidth]{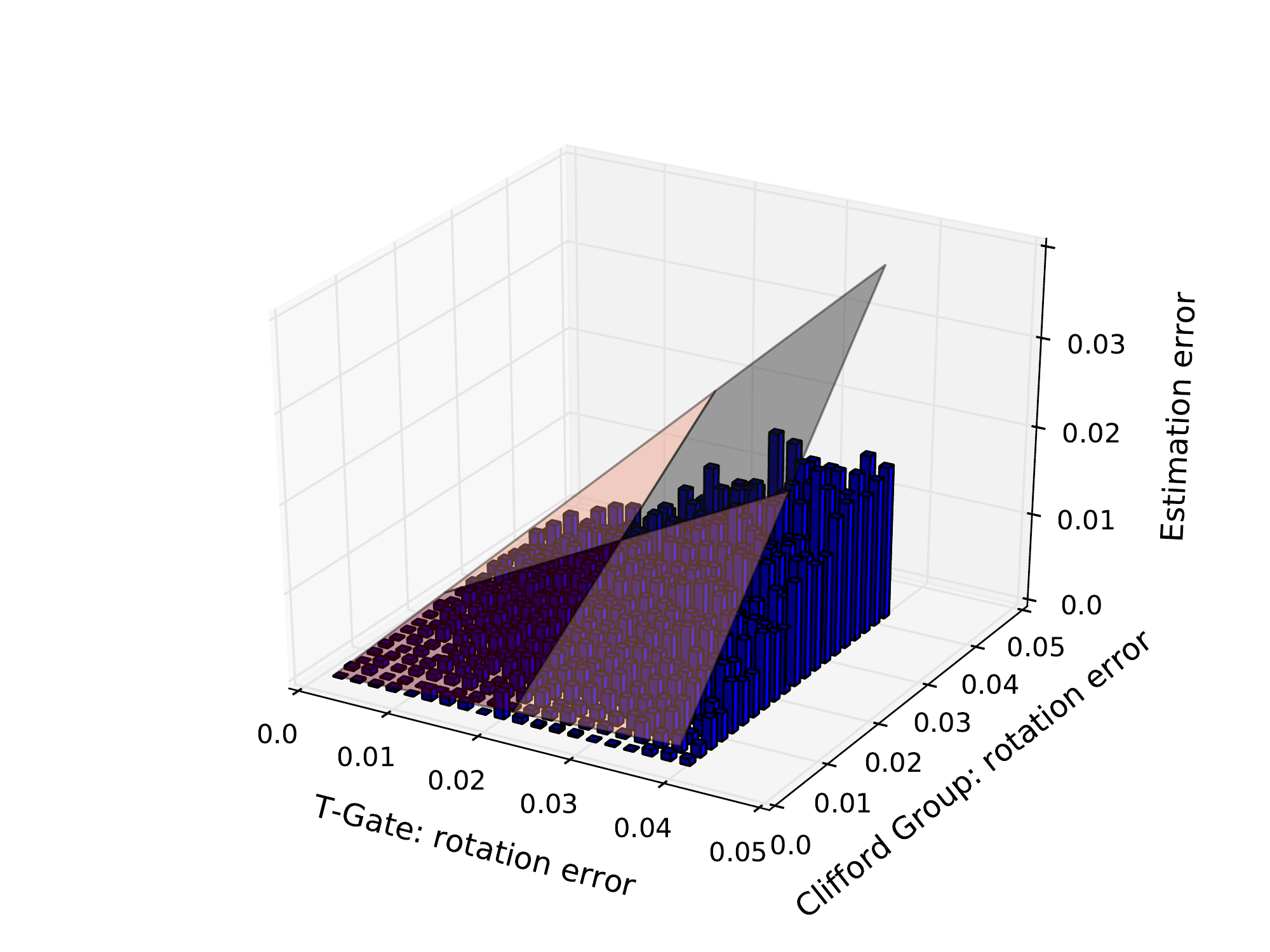}
\caption{Rotation error}
 \label{fig:4}
 \end{subfigure}
\begin{subfigure}[t]{\columnwidth}
 \includegraphics[width=1.1\columnwidth]{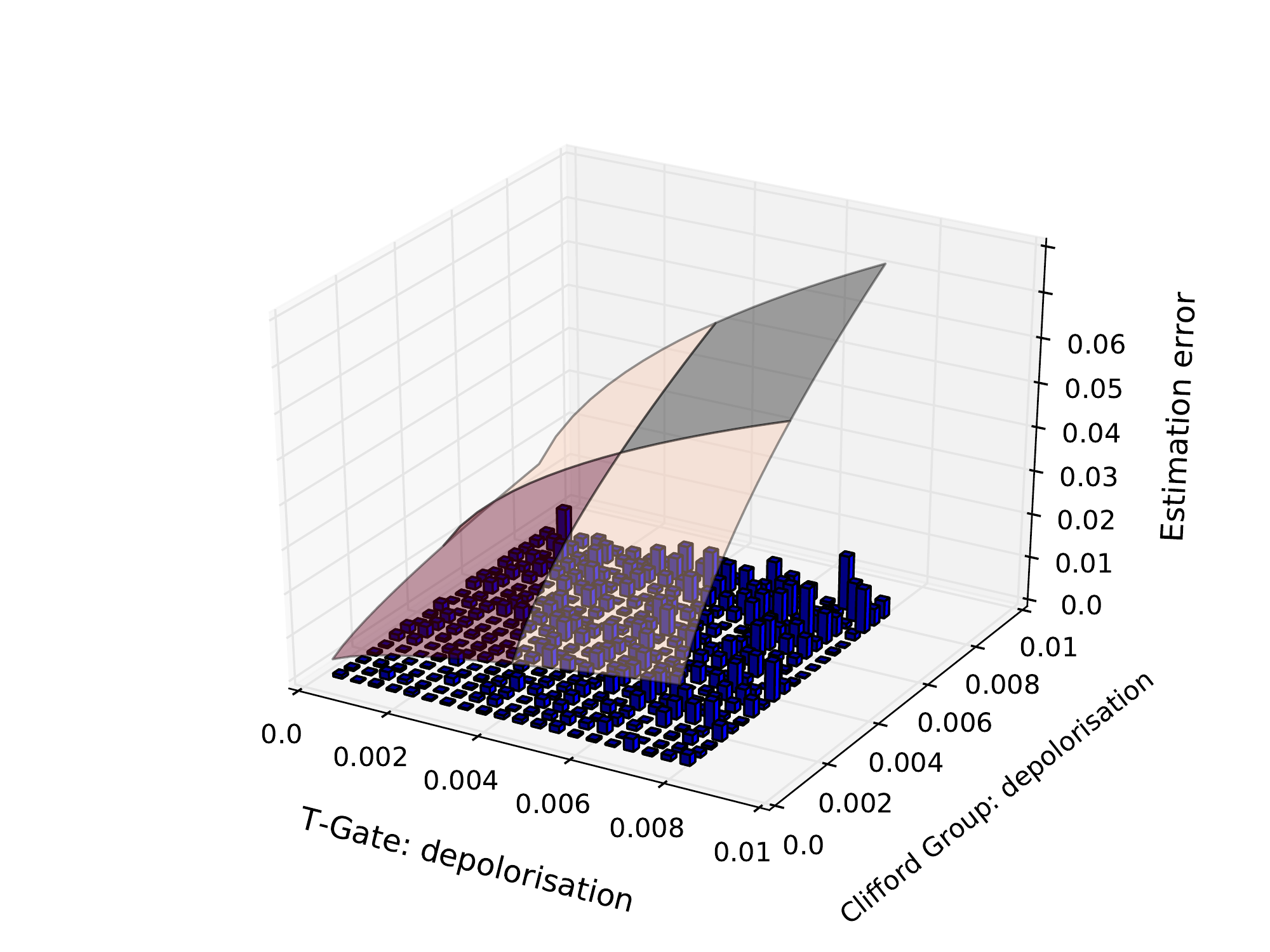}
 \caption{Depolarizing noise}
 \label{fig:5}
\end{subfigure}
\caption{(a) Error in the $T$~gate estimation from various numerical simulations where we have varied the over-rotation levels in the Clifford group and in the $T$~gates (rotation ($\theta$) shown as radians around the x-axis ie $e^{(-iX\frac{\theta}{2})}$). 
The transparent plane above the graph shows the bounds in \autoref{Eq:error}. 
As can be seen the estimate error does not appear to saturate these bounds as the error rate in the Clifford group increases. 
(b) Error in the $T$~gate estimation for a general amplitude dampening channel. 
Each bar represents the error in the estimate for a single run of the protocol using the amplitude dampening $\gamma$ given by the $x,y$ coordinates. 
The transparent plane above the graph shows the bounds calculated in accordance with \autoref{Eq:error}.\hspace*{\fill}}
\end{figure*}

For \autoref{fig:depol} we used a depolarization channel \autoref{eq:dep} composed with the x-axis over-rotation channel ($e^{(-iX\frac{\theta}{2})}$)(see caption for details). 
In this case the fidelity of the Clifford gates is 99.5\% and the $T$~gate 99.0\%. 
The estimate in this case was 98.9\%. 
As can be seen from the the second boxplot in \autoref{fig:2} bounds in \autoref{Eq:error} are not saturated, the standard deviation of the estimated fidelity (over 100 simulations) being 0.001 (0.1\%).
The final illustrative run \autoref{fig:damp} uses an amplitude damping channel parameterized by $p$ and $\gamma$ with the Kraus operators given by 
\begin{align} 
\label{eq:amp}
 E_0&=\sqrt{p}
 \begin{pmatrix}
 1&0\\0&\sqrt{1-\gamma}
 \end{pmatrix}
  & E_1&=\sqrt{p} 
  \begin{pmatrix}
  0&\sqrt{\gamma}\\0&0
  \end{pmatrix}\\
 E_2&=\sqrt{1-p}
 \begin{pmatrix}
 \sqrt{1-\gamma}&0\\0&1
 \end{pmatrix}
  & E_3&=\sqrt{1-p}
  \begin{pmatrix}0&0\\ \sqrt{\gamma}&0
  \end{pmatrix}\nonumber\,.
\end{align}
Choosing $p=0.995$ and $\gamma=0.01$ for the Clifford gates and $p=0.99$ and $\gamma=0.04$ for the $T$~gates, gives us a Clifford gate fidelity of 99.67\% and a $T$~gate actual fidelity of 98.7\% compared with the estimated fidelity of 98.6\%. 
The standard deviation of the estimate over 100 runs was 0.001 (0.1\%).

\autoref{fig:4} and \autoref{fig:5} show the error in the estimate of the $T$~gate over more varied parameters. 
In \autoref{fig:4} the rotational errors on the Clifford gates and $T$~gate were varied. 
On the same graph we have plotted the plane corresponding to the error bounds detailed in \autoref{Eq:error}. 
As can be seen the error bounds are not saturated in the regime of higher rotational errors. 
\autoref{fig:5} shows a similar treatment for the generalized amplitude dampening channel with varying $\gamma$. 
Again we see the protocol performs better than the limits in \autoref{Eq:error}.

Finally we tested the protocol against randomly generated unital CTPT error channels, close to Identity. To create such channels we used the parameterisation of the unital block $\bs{E}_u$ (see \autoref{sec:fidelity}) introduced in \cite{King2001}. The block can be represented as $U\Sigma V$, where $U$ and $V$ are unitary and $\Sigma$ is is diagonal with real entries ($\lambda_1,\lambda_2,\lambda_3$), constrained such that $|\lambda_1 \pm \lambda_2| \leq |1\pm \lambda_3|$. The unitary matrices were generated by combining three rotations (around the x, y and x axis), where the rotation amount was drawn from a normal distribution with $\mu=0$ and a standard a deviation ($\sigma$) of 0.02 (0.1 for the $T$~gate error channel). The diagonal matrix consisted of values drawn from a normal distribution with $\mu=0.98$ ($\mu=0.95 $ for the $T$~gate error channel) and $\sigma=0.01$ ($\sigma=0.1$ for the $T$~gate). The diagonal elements were re-chosen if they exceeded one or failed to satisfy the requirement stated above. As can be seen from the inset to \autoref{fig:channels} this led to random ``Clifford" error channels that were roughly normally distributed between fidelity ranges from 0.98 to 1.0 and $T$~gate error channels with fidelities from 0.9 to 1.0.

Each of these maps was then used as an error channel on the Clifford Gates and the $T$~gate respectively and 1) the fidelity  of the Clifford gates were estimated using a standard Randomized Benchmarking protocol; 2) the fidelity of the Clifford gates were estimated using the Clifford/Pauli protocol introduced in this paper; and 3) the interleaved protocol discussed above was used to estimate the fidelity of the $T$~gate. In each case sequences of 2 to 100 gate lengths (increasing by 4) were simulated, 4000 such random sequences for each `length', with each sequence being run once, returning either a 1 or 0.  As can be seen from \autoref{fig:channels} the ``Clifford Pauli'' twirl protocol used in this paper returned accurate results in the regime of high-fidelity gates and the $T$~gate estimation is accurate to within the theoretical bounds, even where the noise channel for the $T$~gates is randomly different from the noise channel for the Clifford group.

\begin{figure}

  \begin{tikzpicture}
      \node[anchor=south west,inner sep=0] (image) at (0,0) {\includegraphics[width=\columnwidth]{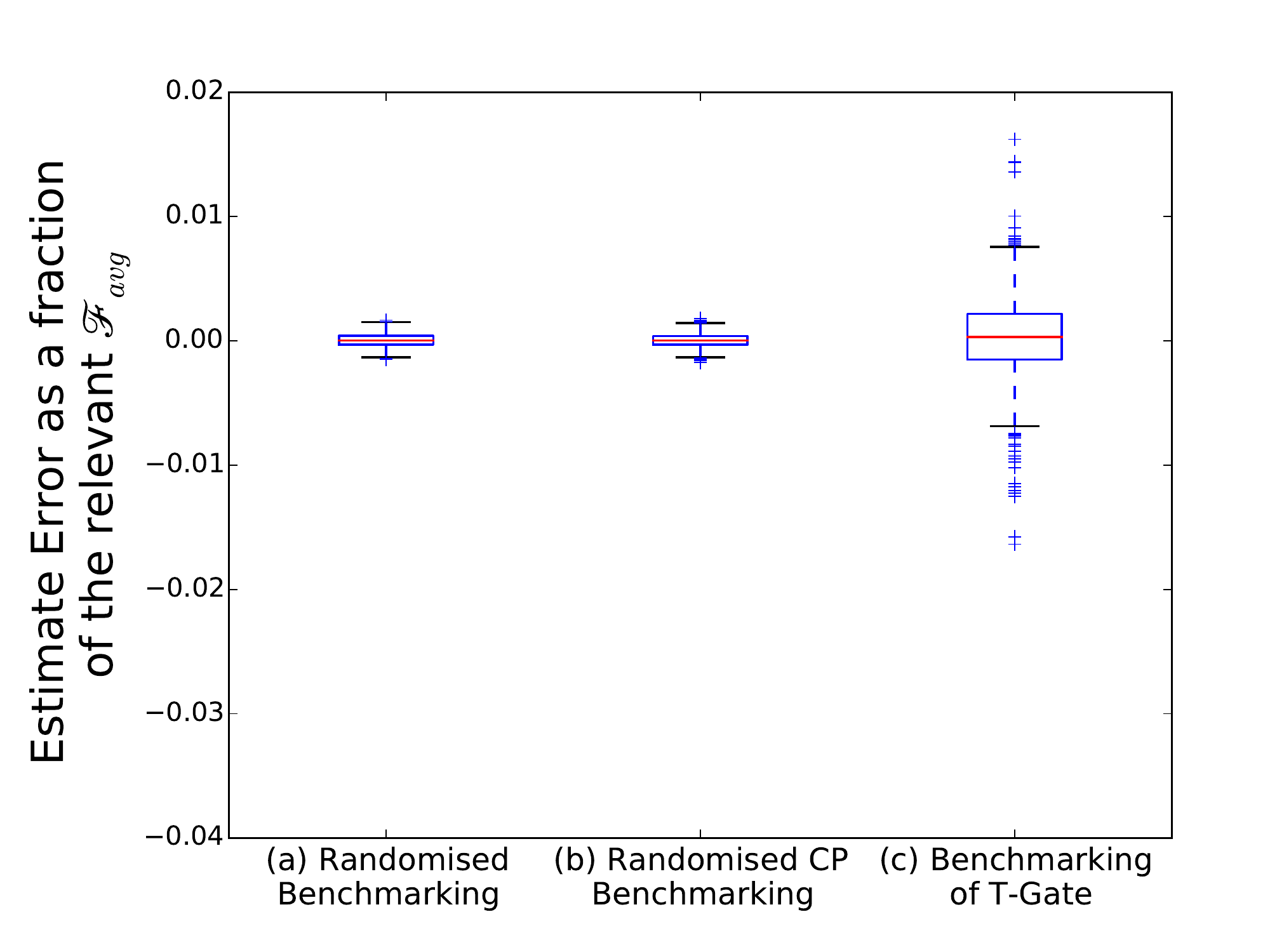}};
       \begin{scope}[x={(image.south east)},y={(image.north west)}]
          \node[anchor=south west,inner sep=0] (image) at (0.2,0.12) {\includegraphics[width=0.5\columnwidth]{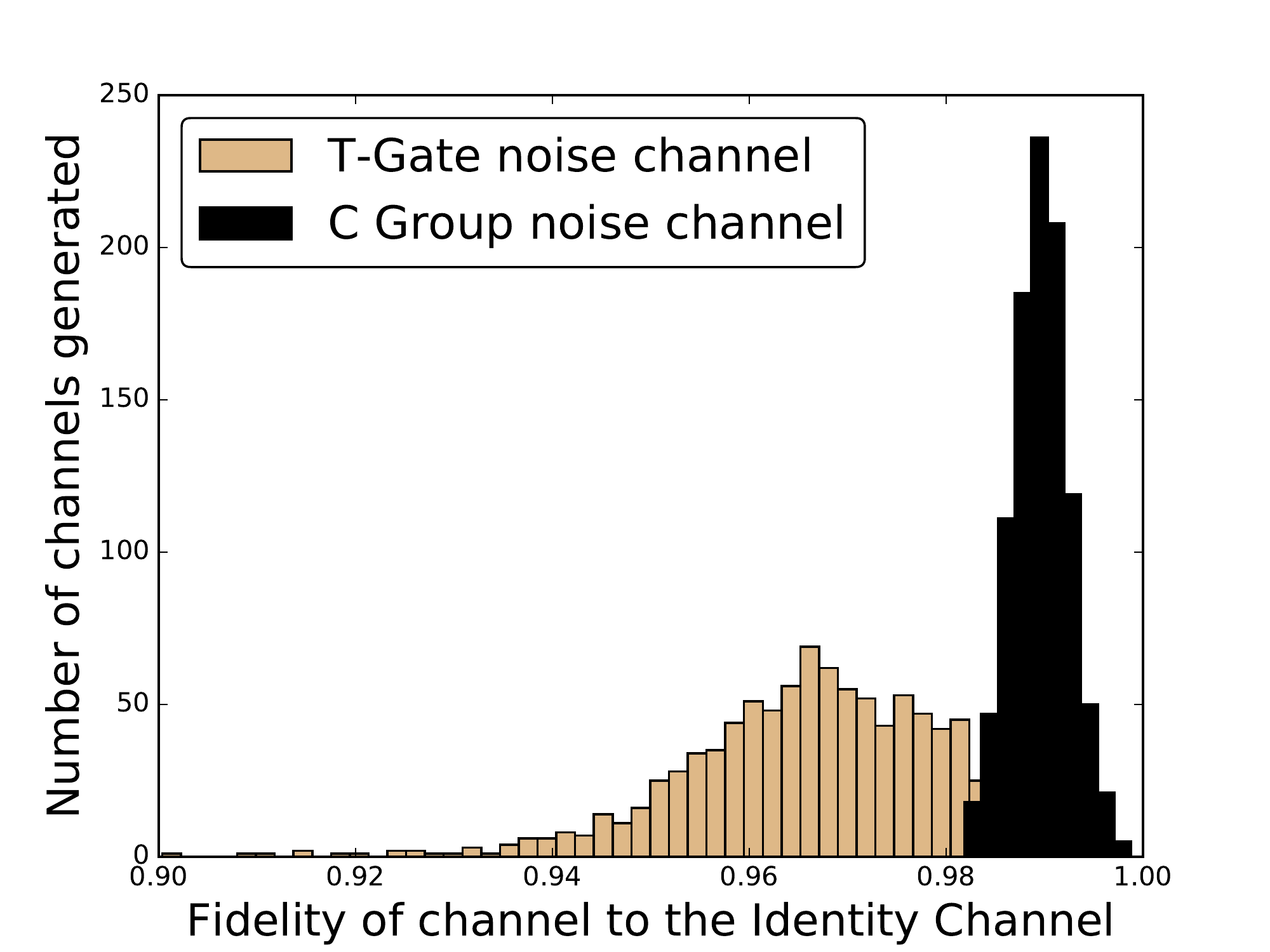}};
       \end{scope}
   \end{tikzpicture}

 \caption{
 Tukey plots showing estimated fidelities over 1000 randomly generated noise channels, taking single measurements at various gate lengths from 4,000 random single-shot sequences per gate length. The noise channels were generated to be near Identity (see text for details) with the noise channel applied to the Clifford group gates having a fidelity to the Identity channel of between 0.98 and 1. The noise channel applied to the $T$~gates had a fidelity to the Identity of between 0.9 and 1. The inset barchart shows the fidelity distribution of these random noise channels. The Tukey plots show the estimation error as a fraction of the actual fidelity (i.e. $(\mathcal{F}_{\text{estimate}}-\mathcal{F}_{\text{actual}})/\mathcal{F}_{\text{actual}}$). Plot (a) shows the results from a ``standard'' Randomized Benchmarking simulation, Plot (b) the reference fidelity generated by the protocol presented in this paper (ie a random Clifford followed by a random Pauli). They are indistinguishable in this regime. The error in the estimation of the $T$~gate is shown and can be further reduced by increasing the number of sequences measured. \hspace*{\fill}
} 
\label{fig:channels}
\end{figure}

\section{Compatibility with small subgroups}\label{sec:cleve} 

In this section we discuss the compatibility of this protocol with the work of \citet{cleve2015}. 
In \cite{cleve2015} it is shown how a subset of the Clifford group that can be used to construct exact unitary-2 designs with a gate complexity that is nearly linear in the number of qubits (in the simplest case it scales as $\mc{O}(n\text{ log }n \text{ log log } n)$ assuming the extended Riemann Hypothesis is true). 
For a given number of qubits $n$, they note that a collection of Clifford gates isomorphic to $\text{SL}_2(\text{GF}(2^n))$ (of which there are $2^{3n}-2^{n}$ gates) when mixed with the Paulis ($2^n$), give a unitary-2 design. 
The mixed gates are a subgroup of Clifford gates, where the total number of gates is $2^{5n}-2^{3n}$. 
 By way of illustration the Clifford group has cardinality $|\mathcal{C}_n|=2^{n^2+2n}\prod_{j=1}^n(4^j-1)$. 
With two qubits this means that we can reduce the 11,520 Cliffords to a mere 60 gates, twirled by the 16 Pauli gates (giving 960 gates in total). 
The main contribution of \cite{cleve2015} is to show how these gates can be constructed with near-linear gate costs (the number of one- and two-qubit gates). 
Here we use their results to reduce the variety of gates needed to conduct randomized benchmarking of $T$~gates. 
Since our changes are only in respect to the additional single-qubit gates, the near-linear time complexity is preserved.

Let $\mathcal{C_\gamma}$ represent the set of gates isomorphic to $\text{SL}_2(\text{GF}(2^n))$. 
As before $\mathcal{P}$ represents the $2^n$ Paulis and $\mathcal{C}_t$ represents the Cliffords formed by conjugation of the elements of $\mathcal{P}$ with a $T$~gate. 
Let $S$ be the phase gate (i.e.\ the gate formed by two applications of a $T$~gate).

The protocol previously discussed needs to be changed so that the reference run consists of n-sequences of gates drawn randomly from ${\mathcal{C}_\gamma}\mathcal{P}$. \cite{cleve2015} contains the details how to generate $\mathcal{C_\gamma}$, the set of gates ${\mathcal{C}_\gamma}\mathcal{P}$, is the unitary 2-design we wish to use and it becomes the group $\mc{G}$ referred to in the protocol detail in \autoref{sec:protocol}.

The interleaved sequence would be physically implemented using repeated applications of the following gates: $\mathcal{C}_{\gamma\prime}T\mathcal{C}_tT$, where $\mathcal{C}_{\gamma\prime}$ is drawn from the gates formed by the sequence $\mathcal{C}_\gamma(S)^\dagger$. That is the $\mc{G}$ in the interleaved part of the protocol (\autoref{sec:interleaved}) consists of the  group formed by the gates ${\mathcal{C}_\gamma}\mathcal{P}(S)^\dagger$ (which are all Clifford gates), rather than the group formed by  ${\mathcal{C}_\gamma}\mathcal{P}$.
As is readily apparent the interleaved sequence can be analysed as:
\begin{equation}
\mathcal{C}_{\gamma\prime}T\mathcal{C}_tT\rightarrow\mathcal{C}_\gamma(S)^\dagger TT\mathcal{P}\rightarrow\mathcal{C}_\gamma\mathcal{P}
\end{equation}
which is an efficiently invertible unitary-2 design as required.

\section{Discussion}
We have provided a simple extension of the randomized benchmarking protocol for interleaved gates that allows the fidelity of $T$~gates to be robustly estimated. 
Unlike previous proposals for benchmarking the $T$~gate our proposal retains the benefits of an exact unitary 2-design that at a minimum at least diagonalizes the noise between the $T$~gates. 
Our numerical analysis confirms that, subject to the usual randomized benchmarking assumptions of time- and gate-independent noise, the fidelity of $T$~gates can be estimated to a high degree of accuracy. 
As discussed above, the analyses in \cite{Epstein2014,Magesan2011,Wallman2014a} will apply, confirming that the gate independence assumption can be dropped without a significant effect on the results. 
Our simulations confirm that the theoretical bounds for interleaved benchmarking are only tight in the region of high gate fidelity where the gate of interest has a worse fidelity than those forming the rest of the benchmarking group. 
Outside such regions the protocol appears to still give accurate estimations of the fidelity of the gate in question. 

Finally we note that the use of the protocol consisting of sequences of a  random Clifford gate followed by a random Pauli gate creates a depolarising channel  that differs from the usual Randomized Benchmarking depolarising channel by the variance of the trace of the diagonal elements $\text{Tr}(P_j\mc{E}(P_j))$ where $P_j$ runs over each of the three traceless Paulis. This provides a method of using Randomized Benchmarking protocols to analyse the anisotropy of the noise, which may provide useful future work.

\begin{acknowledgments}
This work was supported by the US Army Research Office grant numbers W911NF-14-1-0098 and W911NF-14-1-0103, and by the Australian Research Council Centre of Excellence for Engineered Quantum Systems  CE110001013. 
STF acknowledges support from an Australian Research Council Future Fellowship FT130101744. 

The authors would like to thank David Gross, Joel Wallman and Arnaud Dugas for helpful and illuminating discussions.
\end{acknowledgments}

\bibliography{library}

\end{document}